\documentclass[aps,pre,twocolumn,groupedaddress,showpacs]{revtex4}
\usepackage{graphicx}
\usepackage{amsmath}
\begin{document}
\title{Contour dynamics model for electric discharges}
\author{M. Array\'as$^{1}$, M. A. Fontelos$^{2}$ and C. Jim\'enez$^{1}$}
\affiliation{$^{1}$\'Area de Electromagnetismo, Universidad Rey Juan
Carlos, Camino del Molino s/n, 28943 Fuenlabrada, Madrid, Spain}
\affiliation{$^{2}$Instituto de Ciencias Matem\'{a}ticas
(CSIC-UAM-UCM-UC3M), C/ Serrano 123, 28006 Madrid, Spain}

\begin{abstract}
We present an effective contour model for electrical discharges deduced as the asymptotic limit of the minimal streamer model for the propagation of electric discharges, in the limit of small electron diffusion. The incorporation of curvature effects to the velocity propagation and not to the boundary conditions is a new feature and makes it different from the classical Laplacian growth models. For the first time, the dispersion relation for a non planar 2-D discharge is calculated. The development and propagation of finger-like patterns are studied and their main features quantified.  
\end{abstract}

\date{\today}
\maketitle

The electrical breakdown of various media is usually caused by the
appearance and propagation of ionization waves and, in particular,
streamers. The difficulties for the understanding of this important
and interesting phenomena are both experimental and theoretical. From
experimental point of view, the historical lack of quantitative data,
from small to large scales discharges, are mainly due to the fast time
scales of the processes involved. Recently this situation is starting
to change with the development of new experimental techniques. On the
other hand, the theory is still far from being complete. Even in the
simplest hydrodynamic approximation for describing the phenomena, it
is necessary to solve a nonlinear system of balance equations,
together with Poisson equation, which poses a challenging problem both
numerical and analytical. Among the
recent progress in the understanding of the propagation mechanism we
can mention: the study of stationary plane ionization waves
\cite{r1,vanS}, first self-similar solutions for ionization waves in
cylindrical and spherical geometries \cite{aft,r3}, and the formation
of streamers as the result of the instability of planar ionization
fronts \cite{ME,aftprl,abft}. In the simplest hydrodynamic
approximation, the fronts are subject to both stabilizing forces due
to diffusion which tend to dampen out any disturbances, and
destabilizing forces due to electric field which promote them.

In this letter we introduce and justify theoretically a contour
dynamical model which allows us to make some progress in 2-D and 3-D
more general situations. Here we will use ideas coming from the
context of electrohydrodynamics. More precisely, we will arrive at a
model similar to the celebrated Taylor-Melcher leaky dielectric model
for electrolyte solutions \cite{S}, but suitably adapted to the
context of electric (plasma) discharges. We will obtain the dispersion
relation for 2-D discharges like those described in some experiments \cite{Japos}.

For our contour model, in the asymptotic limit of small diffusion $D\ll 1$
(electron diffusions are typically of order $0.1$ m$^2$/s and, since typical velocities and streamer size [1] are $10^{5}$ m/2 and $10^{-2}$ m we get $D$ of order $10^{-3}$), the interface will move according to
\begin{equation}
v_{N}=-E_{\nu }^{+}+2\sqrt{D\alpha(E_{\nu }^{+})} -D \kappa
\label{bb1}
\end{equation}%
being $v_N$ the normal velocity and $\kappa$ twice the mean curvature and $\alpha(E_{\nu }^{+})$ to be defined below. The free charge density on the boundary will be concentrated  in a diffuse layer, and for the limit considered, its behavior will be characterized by a negative charge surface density $\sigma$ given by
\begin{equation}
\frac{\partial \sigma }{\partial t}+\kappa v_{N}\sigma =-\frac{E_{\nu
}^{-}}{\varrho}-j_\nu^{-},
\label{bb4}
\end{equation}%
where $j_\nu^{-}$ is the current density coming from the ionized region ($\Omega$) to its boundary ($\partial \Omega$) in the normal direction and $\varrho$ is related to the resistivity of the plasma. $E_{\nu }^{\pm}$ is the normal component of the electric field at the
interface when approaching it from the region without/with plasma, i.e %
\begin{equation}
E_{\nu }^{+}=\lim_{\mathbf{x}%
\rightarrow (\partial \Omega )^{+}}\left( -\frac{\partial V}{\partial \nu }%
\right) \ , \ E_{\nu}^{-}=\lim_{\mathbf{x}\rightarrow (\partial \Omega )^{-}}\left( -%
\frac{\partial V}{\partial \nu }\right) ,
\label{bb2}
\end{equation}
with $V$ found by solving%
\begin{equation}
\Delta V=\sigma\, \delta(\mathbf{r}-\mathbf{r}_{\partial \Omega }).
\label{lapv}
\end{equation}%
Notice that in the case $\varrho^{-1}\gg 1$, we arrive to Lozansky-Firsov
model \cite{LF} with a correction due to electron diffusion, meanwhile in
the limit $D = 0$ we arrive at the classical Hele-Shaw model. Such a
model is known to posses solutions that develop singularities in the
form of cusps in finite time \cite{PK} but, when regularized by
surface tension corrections, the interface
may develop various patterns including some of fractal-type (see
\cite{Low} for a recent development and references therein).

Let's derive the equations \eqref{bb1} and \eqref{bb4}. We start from a minimal
model of the discharge that in dimensionless units reads \cite{ME}
\begin{eqnarray}
\frac{\partial n_{e}}{\partial t}-\nabla \cdot \left( n_{e} {\mbox{\bf E}} +
D \, \nabla n_{e} \right) &=& n_{e} \alpha(|\mathbf{E}|),  \label{elec1} \\
\frac{\partial n_{p}}{\partial t} &=& n_{e} \alpha(|\mathbf{E}|),  \label{ion1} \\
\nabla \cdot \mathbf{E} &=& n_{p} - n_{e},  \label{gauss1}
\end{eqnarray}
with $\alpha(|\mathbf{E}|)=|\mathbf{E}|\exp(-1/|\mathbf{E}|)$.  Next we derive a
contour dynamics equation for the evolution of the interface between
the region occupied by the plasma and the region free (or with a very
small density) of plasma. Note that this interface does not exist as
such but, as in the case of planar fronts, the region where $n_{e}$
jumps from values close to $1$ to values close to $0$ has a thickness
$O(D^{\frac{1}{2}})$ \cite{aft}. Here we will give an equation, valid
in the asymptotic limit $D\ll 1$.

We take a level surface of $n_e$ representing the interface, and
introduce local coordinates $\tau $ (along the level surface of $n_{e}$) and $\nu$ (orthogonal to the level surface of $n_{e}$). Since the front thickness
is $O(D^{\frac{1}{2}})$ it is natural to introduce $ \nu
=D^{\frac{1}{2}}\chi $, so we can write approximately
\[
D\,\Delta =\frac{\partial ^{2}}{\partial \chi ^{2}}+\sqrt{D}\kappa \frac{%
\partial }{\partial \chi }+D\left( \Delta _{\perp}-\kappa ^{2}\chi \frac{%
\partial }{\partial \chi }\right) +O(D^{\frac{3}{2}})
\]%
where $\Delta _{\perp}$ is the transverse laplacian and $\kappa$ is twice the mean curvature in 3-D or just the curvature in 2-D (see e.g. \cite{Pismen}).
Hence from \eqref{elec1} and \eqref{gauss1}
\[\begin{split}
\frac{\partial n_{e}}{\partial t}-E_{\tau }\frac{\partial n_{e}}{\partial
\tau }-\left( \frac{E_{\nu }}{\sqrt{D}}+\sqrt{D}\kappa \right) \frac{%
\partial n_{e}}{\partial \chi }-\frac{\partial ^{2}n_{e}}{\partial \chi ^{2}}%
=\\ = n_{e}\left(\alpha(|\mathbf{E}|)+n_{p}-n_{e}\right) +O(D).
\end{split}
\]%

We expand now $|\mathbf{E}|$ and $n_{p}$ at the diffusive boundary layer as
explained in \cite{aft} and \cite{abft}, so it is straightforward to end with
\begin{equation}
\frac{\partial n_{e}}{\partial t}-E_{\tau }\frac{\partial n_{e}}{\partial
\tau }-\left( \frac{E_{\nu }}{\sqrt{D}}-2\sqrt{\alpha(E_{\nu })}+\sqrt{D}\kappa \right) \frac{%
\partial n_{e}}{\partial \chi }=O(D^{\frac{1}{2}}),  \label{transne}
\end{equation}%
where a curvature term appears and is relevant provided $ 1\ll \kappa \ll D^{-\frac{1}{2}}$.
Eq.~(\ref{transne}) is a transport equation for the electron density
so that the level line of $n_{e}$ that we have chosen to describe the
evolution of the interface moves with a normal velocity given by \eqref{bb1}.
Notice that the tangential velocity does not change the geometry of the
curve during its evolution. Nevertheless, tangential exchanges of charge
affect the evolution through the dependence of $v_{N}$ on $E_{\nu }$.

Next we describe the charge transport along the interface. We trace a ``pillbox'' $\mathcal{D} $ around a surface element. The region $%
\mathcal{D} $ will be such that $\Delta \tau \gg \Delta \nu $ and will contain
both the diffusive layer for $n_{e}$ and the region where $n_{e}-n_{p}$ (the
net negative charge) has significant values. If we subtract \eqref{ion1} from \eqref{elec1} and integrate over $\mathcal{D} $ we find
\begin{equation}
\begin{split}
\frac{\partial}{\partial t}\int_{\mathcal{D} }(n_{e}-n_{p})\,dV&=\int_{\partial
\mathcal{D} }\left( n_{e}{\mbox{\bf E}}+D\,\nabla n_{e}\right) \cdot \mathbf{n}dS
=\\ & = \left. n_{e}E_{\nu }\Delta \tau \right]_{\chi = -\infty }^{\infty }+O(D^{\frac{1}{2}}),
\end{split}
\label{s1}
\end{equation}
where we have used that $n_{e}\rightarrow 0$ for $\chi =\nu /\sqrt{D}\gg 1$,
$\left\vert \nabla n_{e}\right\vert \rightarrow 0$ for $\left\vert \chi
\right\vert \gg 1$ and we have neglected the contributions of the lateral
transport of charge by $E_{\tau }$ in comparison with exchange of charge
with the bulk by $E_{\nu }$. This assumption is also common in the
Taylor-Melcher model and we will follow it.
Notice that
\begin{equation}
\frac{\partial }{\partial t}\int_{\mathcal{D} }(n_{e}-n_{p})\,dV=\frac{\partial }{%
\partial t}\int_{-\infty }^{\infty }(n_{e}-n_{p})\Delta \tau d\nu =\frac{%
\partial }{\partial t}\left( \sigma \Delta \tau \right),
\label{s2}
\end{equation}
and the length (or area) element of the surface will suffer a change during
interface deformation given by%
\begin{equation}
\frac{\partial \Delta \tau }{\partial t}=\kappa v_{N}\Delta \tau.
\label{s3}
\end{equation}
We can also include a source (an insulated wire inside the plasma, for instance at $\mathbf{x}_0$, carrying a electric current $I(t)$). This source will create a current density inside the plasma and as quasineutrality is fulfilled, we will have for the interior of $\Omega$
\begin{equation}
\nabla \cdot \mathbf{j} = I(t)\delta(\mathbf{x}-\mathbf{x}_0).
\label{nablaj}
\end{equation}
By putting \eqref{s1}-\eqref{nablaj} together we can finally write the equation \eqref{bb4},
%\begin{equation}
%\frac{\partial \sigma }{\partial t}+\kappa v_{N}\sigma =-\frac{E_{\nu }^{-}}{\varrho}-j_\nu^{-}\,,
%\label{sigmaeq}
%\end{equation}%
where $\varrho$ is the effective resistivity of the medium, given in this case as $\varrho^{-1}=\lim_{\chi =-\infty }n_{e}$ and $E_{\nu }^{-}$ is the normal component of the electric field when approaching the interface from inside the plasma
region. $j_\nu^{-}$ is normal component of the current density given by \eqref{nablaj} arriving at the interface. There is a jump in the normal component of the electric field across the interface given by
\begin{equation}
E_{\nu }^{+}-E_{\nu }^{-}=-\sigma.
\label{jump}
\end{equation}%
Eq.~(\ref{bb4}) will provide the surface charge density $\sigma $
as a function of time. From it, we can compute the electric field and move
the interface with (\ref{bb1}). Two limits can be easily identified in the case of $\mathbf{j}\approx 0$: a) the
limit of large conductivity%
\[
\varrho^{-1}\gg 1,\ E_{\nu }^{-}=0\Rightarrow \ V=C(t).
\]%
so that the interface is equipotential and b) the limit of small
conductivity%
\[
\varrho^{-1}\ll 1,\ \frac{\partial }{\partial t}\left( \sigma \Delta \tau \right)
=0\Rightarrow \sigma \Delta \tau =Const.
\]%
where the charge contained by a surface element is constant and the density
only changes through deformation (with change of area) of the interface.

\begin{figure}
\centering
\includegraphics[width=0.5\textwidth]{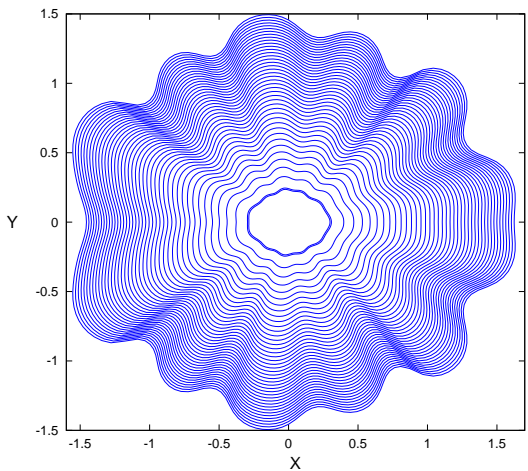}
\caption{The evolution of an initial arbitrary front with a constant net charge $Q=30$ and $D=0.1$.}
\label{fig1}
\end{figure}

Next we study as an application a 2-D case. The 2-D case is not at all academic as there have been experiments where a 2 dimensional streamer discharge is created on a surface and branched pattern are observed.
% To our knowledge, these experiments are one of the few ones in which quantitative data is available about the electric field a charge distribution in the discharge, so the theory may be tested. We suppose that like in the experiment, a 2-D discharge is created by applying a potential difference to a wire at the center of a BGO crystal ($Bi_4Ge_3O_{12}$, with relative permittivity equals 16). One side of the BGO has a transparent conductive coating and is grounded. The reverse side has a dielectric mirror coating. The BGO has Pockels effect and the polarization of the light transmitted through it allows to measure the 2-D air dischare tested. We suppose that like in the experiment, a 2-D discharge is created by applying a potential difference to a wire at the center of a BGO crystal ($Bi_4Ge_3O_{12}$, with relative permittivity equals 16). One side of the BGO has a transparent conductive coating and is grounded. The reverse side has a dielectric mirror coating. The BGO has Pockels effect and the polarization of the light transmitted through it allows to measure the 2-D air discharge on the dielectric mirror side (see \cge on the dielectric mirror side (see \cite{Japos} for details).
In Fig.~\ref{fig1} we can see a numerical simulation intended to mimic the experiment of \cite{Japos} using \eqref{bb1} and \eqref{bb4}. This set of equations also allow us to calculate the dispersion relation for the front stability and give some analytical insight.

Let's write the position and the charge ``surface'' density of the interface as
\begin{eqnarray}
  \label{pert1}
  r(\theta,t)=r(t)+\delta S(\theta,t)\\
  \label{pert2}
  \sigma(\theta,t)=-\frac{Q(t)}{2\pi r(\theta,t)}+\delta \Sigma(\theta,t)
\end{eqnarray}
where $r(t)$ is the solution of the equations for the radial symmetrical front, $Q(t)= \int_0^t I(t)\,dt$ and $\delta$ is our bookkeeping small parameter for the expansions that follow.

We will start calculating the correction to the electric field due to the geometrical perturbation of the surface and the extra charge deposited on it. The electric potential will be up to delta order
$\Phi(\mathbf{x})=V(\mathbf{x})+\delta V_p(\mathbf{x})$
being $V$ the solution for the symmetrical problem. The term  $V_p(\mathbf{x})$ will satisfy the equation $\Delta V_p = O(\delta)$
and further on, we will change coordinates to $\mathbf{x} \longrightarrow \mathbf{\tilde{x}}=\mathbf{x}\, r(t)/r(\theta,t)$
so the perturbed surface becomes a disk of radius $r(t)$. Now at zero order $V_p$ satisfies the Laplace equation with the boundary being a disk. Hence we have in the new polar coordinates
\begin{equation}
V_p(\tilde{r},\theta)=\begin{cases}
    \sum_1^{\infty}\psi_n \cos(n\theta)\left(\frac{r}{\tilde{r}}\right)^n,& \tilde{r} >  r\\
    \sum_1^{\infty}\varphi_n \cos(n\theta)\left(\frac{\tilde{r}}{r}\right)^n, & \tilde{r} \le r
  \end{cases}
\label{cv1}
\end{equation}
where we have imposed the conditions of $V_p$ remains finite at the origin and becomes zero at very large distances. The potential at the surface position coming from the exterior is $\Phi(\mathbf{x}_s^+) = - Q(t)\log(r(\theta,t))/2 \pi +\delta V_p(\mathbf{x}_s^+)$
and from the interior is $\Phi(\mathbf{x}_s^-) = C(r(t),t)+\delta V_p(\mathbf{x}_s^-)$
where $C(r(t),t)$ is a function independent of $\theta$.
So if we write the surface perturbation as
\begin{equation}
S = \sum_{n=1}^{\infty}s_n(t)\cos(n\theta),
\label{sn}
\end{equation}
 imposing the condition of continuity for the potential,
we find that the coefficients of the potential in \eqref{cv1} are related by
\begin{equation}
\psi_n = \varphi_n + \frac{Q(t)}{2\pi r} s_n.
\label{coeff}
\end{equation}
Now we can calculate the electric field to $\delta$ order. Changing back coordinates, from $\mathbf{\tilde{x}}$ to $\mathbf{x}$, the normal components of the electric field at both sides of the surface are
 \begin{eqnarray*}
E_{\nu}^+&=&\frac{Q(t)}{2\pi (r+\delta S)}+\delta\sum_1^{\infty}\left(\varphi_n + \frac{Q(t)}{2\pi r}s_n\right)\frac{n}{r}\cos(n\theta), \\
E_{\nu}^-&=&-\delta\sum_1^{\infty}\varphi_n \frac{n}{r}\cos(n\theta).
\end{eqnarray*}
Then, the jump condition \eqref{jump} together with \eqref{pert2} give the following expression for the charge perturbation
\begin{equation}
\Sigma = -\sum_{n=1}^\infty \left(2\varphi_n + \frac{Q(t)}{2\pi r} s_n \right)\frac{n}{r}\cos(n\theta),
\label{sigman}
\end{equation}
where $r = r(t)$ is the evolving radius of the unperturbed circle, i.e. the radially symmetric solution to (\ref{bb1}),(\ref{bb4}).

We still need the expression of the curvature at $\delta$ order to find the dynamics of the front. In polar coordinates, after introducing the perturbation \eqref{pert1}, the curvature turns out to be
\[\kappa = \frac{r^2+2rS\delta-rS_{\theta\theta}\delta+O(\delta^2)}{\left(r^2+2rS_{\theta}\delta+O(\delta^2)\right)^{\frac{3}{2}}} = \frac{1}{r}-\frac{S+S_{\theta\theta}}{r^2}\delta+O(\delta^2),\]
and the normal component of the velocity
\begin{equation}
v_N = \mathbf{v} \cdot \mathbf{n} = \frac{d r(t)}{d t} +\delta \frac{\partial S(\theta,t)}{\partial t}.\nonumber
\label{vp}
\end{equation}
Now Eqs.~\eqref{bb1} and \eqref{bb4} after introducing the perturbations to first order, defining $\varepsilon=D$, and doing some algebraic manipulations yield
\begin{eqnarray}
\frac{ds_n}{dt} &=& -\left[\frac{Q(t)}{2\pi r^2}(n-1)+\frac{\varepsilon}{ r^2}(n^2-1)  \right]s_n -\frac{n}{r}\varphi_n,
\label{esn}\\
\frac{d\varphi_n}{dt} &=& \biggl[\frac{Q(t)}{2\pi r^2}\left(\frac{Q(t)}{2\pi r}+\frac{(n+2)\varepsilon}{2r}-\varepsilon^{\frac{1}{2}}\sqrt{\alpha(|Q|/2\pi r)}\right)\nonumber\\ &\times&(n-1)-\frac{I(t)}{4\pi r}\biggr]s_n+\frac{1}{2}\left(\frac{Q(t)}{2\pi r^2}n-\frac{1}{\varrho}\right)\varphi_n.
\label{evpn}
\end{eqnarray}
Thus the time evolution of each particular mode has been obtained and it is governed by \eqref{esn} and \eqref{evpn}.
First let's study the limit of ideal conductivity. It makes $\varrho \to 0$, so we can see from \eqref{evpn} that $\varphi_n \to 0$. Physically it means that in the limit of very high conductivity, the electric field inside goes to zero ($E_\nu^- \to 0$), as we approach to the behaviour of a perfect conductor. If we consider that $Q(t)=Q_0$ is constant or its variation in time is small compared with the evolution of the modes (which also implies $I(t)\to 0$), and the same for the radius of the front $r(t)=r_0$, we can try a solution $s_n = \exp(\omega_n t), \varphi_n =0$, to our system, and we get a discrete dispersion relation of the form
\begin{equation}
\omega_n=-\frac{Q_0}{2\pi r_0^2}(n-1)-\frac{\varepsilon}{r_0^2}(n^2-1).
\label{disp1}
\end{equation}

Next we consider the limit of finite resistivity, but such that the total charge is constant at the surface, or varies very slowly. Writing \eqref{evpn} as
\[\frac{d\varphi_n}{dt}=-\frac{d}{dt}\left(\frac{Q(t)}{4\pi r}s_n\right)-\frac{Q(t)}{4\pi r^2}\frac{dr}{dt}n s_n-\frac{1}{2\varrho}\varphi_n,\]
we have now
\[\frac{d\varphi_n}{dt}=-\frac{Q_0}{4\pi r_0}\frac{ds_n}{dt}-\frac{1}{2\varrho}\varphi_n.\]
For a small enough conductivity, so no extra charge reaches the surface, we find $\varphi_n =-\frac{Q_0}{4\pi r_0}s_n$, and with $s_n = \exp(\omega_n t)$, \eqref{esn} yields
\begin{equation}
\omega_n=-\frac{Q_0}{2\pi r_0^2}\left(\frac{n}{2}-1\right)-\frac{\varepsilon}{r_0^2}(n^2-1).
\label{disp2}
\end{equation}

In a curved geometry we can see that the modes are discrete.
If we compare \eqref{disp1} and \eqref{disp2}, for $n=1$ corresponds the same $\omega_1=0$ which reveals the translation invariance of the front. However, for small $n$, there is a $1/2$ factor discrepancy in the dispersion curve between the two limits. In \cite{CPRL}, the origin of this prefactor was discussed for planar fronts (the dispersion relation for planar fronts was obtained in the case of constant charge in \cite{abft}). So imposing constant charge at the surface, this factor is independent of the planar or curve geometry. On the other hand, imposing constant potential at the surface gives a factor of $1$. The intermediate situations can be studied by solving the system \eqref{esn} and \eqref{evpn}.
Another consequence is that in both cases the maximum growth correspond to a perturbation with 
\begin{equation}
n \propto|Q_0|/D,
\label{disp3}
\end{equation}
implying that the number of fingers increases with the net charge and decreases with electron diffusion.
\begin{figure}
\centering
\includegraphics[width=0.45\textwidth]{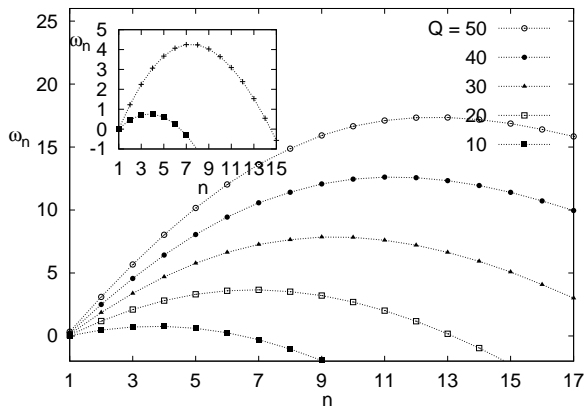}
\caption{Dispersion relation for the discrete modes of a perturbation keeping the charge constant for different charges $Q=-Q_0$. In the inset, the case $Q=10$ is compared with the case of ideal conductivity (cross points).}
\label{fig2}
\end{figure}

To test the analytical predictions, we have calculated numerically the dispersion relation curves for these two cases. In Fig.~\ref{fig2}, for different charges $\omega_1=0$, the slope increases with the amount of charge, the maxima moves to higher modes, and for larger $n$'s the dispersion curves become negative as predicted by \eqref{disp2}. In the inset of Fig.~\ref{fig2} the dispersion relation curve is calculated for the case of constant charge and constant potential. The slope around the origin $n=1$ is bigger for the case of ideal conductivity, when the interface is equipotential. The ratio of the slopes can be estimated as $\approx 0.49$ which is in agreement with our prediction of $1/2$ (see \cite{CPRL}). 

To conclude, we have introduced a contour dynamics model a la
Taylor-Melcher \cite{S} for the streamers discharges. Our model
contains as a particular limit the Lozansky-Firsov model \cite{LF}
with a correction due to electron diffusion which effectively acts as
a surface tension. In the limit $D = 0$ the classical Hele-Shaw model
\cite{PK,Low} is recovered. In this framework we have studied the
stability of 2-D discharges. It is our hope that this model can open
the door for the study of 3-D problems of streamer discharges in
realistic geometries. An extra advantage of this model is its
relatively low computational cost and the high analytical insight
which provides.

The authors thank support from the Spanish Ministerio de
Educaci\'on y Ciencia under projects ESP2007-66542-C04-03, AYA2009-14027-C05-04 and MTM2008-0325.

\end{document}